\documentclass[aps,showpacs]{revtex4}
\usepackage{graphicx}

\def\ket{\rangle}
\def\<{\langle}
\def\>{\rangle}

\begin{document}
\title{A Two-Step Quantum Direct Communication Protocol Using Einstein-Podolsky-Rosen Pair Block}
\author{ Fu- Guo Deng$^{1,2}$, Gui Lu Long$^{1,2,3,4}$\thanks{Corresponding
author:gllong@tsinghua.edu.cn} and Xiao- Shu Liu$^{1,2}$}
\address{
$^1$ Department of Physics, Tsinghua University, Beijing $100084$, P. R.
China\\
$^2$ Key Laboratory For Quantum Information and Measurements,
Beijing 100084
, P. R. China\\
$^3$ Center for Atomic and Molecular NanoSciences, Tsinghua
University, Beijing 100084, P. R. China\\
$^4$ Institute of Theoretical Physics, Chinese Academy of Sciences, Beijing $%
100080$, P. R. China}
\date{\today }

\begin{abstract}
A protocol for quantum secure direct communication using blocks of
EPR pairs is proposed. A set of ordered $N$ EPR pairs is used as a
data block for sending secret message directly. The ordered $N$
EPR set is divided into two particle sequences, a checking
sequence and a message-coding sequence. After transmitting the
checking sequence, the two parties of communication check
eavesdropping by measuring a fraction of particles randomly
chosen, with random choice of two sets of measuring bases. After
insuring the security of the quantum channel , the sender, Alice
encodes the secret message directly on the message-coding sequence
and send them to Bob. By combining the checking and message-coding
sequences together, Bob is able to read out the encoded messages
directly. The scheme is secure because an eavesdropper cannot get
both sequences simultaneously. We also discuss issues in a noisy
channel.

\end{abstract}

\maketitle

\section{INTRODUCTION}
\label{ss1}

The goal of cryptography is to ensure that the secret message is
intelligible only for the two authorized parties of communication
and not be altered during the transmission. Thus far, it is
trusted that the only proven secure crypto-system is the
one-time-pad scheme in which the secret key is as long as the
message.  The two distant parties who want to transmit their
secret message must distribute the secret key first. But it is
difficult to distribute securely the secret key through a
classical channel. Quantum key distribution (QKD), the approach
using quantum mechanics principle for the distribution of secret
key is the only proven protocol for secure key distribution.

A lot of attention has been focused on QKD and it has been
developed quickly since Bennett and Brassard proposed the standard
BB84 QKD protocol\cite{BB84} in 1984. Now there have been a lot of
theoretical
QKD schemes, for instance in Refs.\cite%
{BB84,Ekert,BBM,B92,BW,GV,HIGM,KI,BruB,HKH,CabelloL,CabelloA,LL,SJG,XLG,DLMXL,PBTB,LCA}%
. They can be attributed to one of the two types, the
non-deterministic one and the deterministic one. The feature of
the non-deterministic schemes is that the sender, Alice chooses
randomly two sets of measuring bases (there are at least two sets
of non-orthogonal bases) to produce two kinds of orthogonal states
and transmits them to the receiver, Bob. Bob then also chooses
randomly one of the two sets of bases to measure the states. There
are only a certain probability that Alice and Bob choose the same
bases. So Alice cannot determine which bit value Bob can receive
before they exchange classical information. The typical schemes
are the BB84 \cite{BB84}, Ekert91\cite{Ekert}, BBM92\cite{BBM} and
6-state protocols\cite{BruB}. In contrast, in the deterministic
schemes, Alice and Bob choose the same orthogonal bases for their
measurements, so that they get the same results deterministically
if the quantum channel is not disturbed.
 Typical such protocols are the ones presented in Refs\cite%
{BW,GV,KI,HKH,LL}.

Different from key distribution whose object is to establish a
common random key between two parties, a secure direct
communication is to communicate important messages directly
without first establishing a random key to encrypt them. Thus
secure direct communication is more demanding on the security.
 As a secure direct communication, it must satisfy two
requirements. First the secret messages should  be read out
directly by the legitimate user, Bob when he receives the quantum
states, and no additional classical information is needed after
the transmission of qubits. Secondly the secret messages which
have been encoded already in the quantum states should not leak
even though an eavesdropper may get hold of channel. That is to
say, the eavesdropper can not only be detected  but also obtains
blind results. As classical message can be copied fully, it is
impossible to transmit secret messages directly through classical
channels. But when quantum mechanics enters into the
communication, the story will change.

Recently, Beige et al. proposed a QSDC scheme\cite{BEKW}. In this
scheme the message can be read only after a transmission of an
additional classical information for each qubit. Bostrom and
Felbingeer put forward a Ping-Pong QSDC scheme\cite{BF}. It is
secure for key distribution, quasi-secure for direct secret
communication if perfect quantum channel is used. However it is
insecure if it is operated in a noisy quantum channel, as shown by
W\.{o}jcik\cite{Wojcik}. There is some probability that a part of
the message might be leaked to the eavesdropper, Eve, especially
in a noisy quantum channel. Because Eve can use the
intercept-resending strategy to steal some secret message even
though Alice and Bob will find out her in the end of
communication, especially in a noise quantum channel. Moreover the
capacity is restricted, and an entangled state (an EPR pair ) only
carries one bit of classical information.

In this paper, we will introduce a QSDC scheme with EPR pairs
generalizing the basic ideas in Ref.\cite{LL} in QKD. It will be
shown that it is provably secure and has high capacity. we discuss
the problems in a lossy quantum channel.

\section{THE TWO-STEP QUANTUM SECURE DIRECT COMMUNICATION SCHEME}
\label{ss2}

 An EPR pair can be in one of the four Bell states,

\begin{eqnarray}
\left\vert \psi ^{-}\right\rangle &&=\frac{1}{\sqrt{2}}(\left\vert
0\right\rangle _{A}\left\vert 1\right\rangle _{B}-\left\vert
1\right\rangle _{A}\left\vert 0\right\rangle _{B})  \label{EPR1}\\
\left\vert \psi ^{+}\right\rangle &&=\frac{1}{\sqrt{2}}(\left\vert
0\right\rangle _{A}\left\vert 1\right\rangle _{B}+\left\vert
1\right\rangle _{A}\left\vert 0\right\rangle _{B})  \label{EPR2}\\
\left\vert \phi ^{-}\right\rangle &&=\frac{1}{\sqrt{2}}(\left\vert
0\right\rangle _{A}\left\vert 0\right\rangle _{B}-\left\vert
1\right\rangle _{A}\left\vert 1\right\rangle _{B})  \label{EPR3}\\
\left\vert \phi ^{+}\right\rangle &&=\frac{1}{\sqrt{2}}(\left\vert
0\right\rangle _{A}\left\vert 0\right\rangle _{B}+\left\vert
1\right\rangle _{A}\left\vert 1\right\rangle _{B})  \label{EPR4}
\end{eqnarray}

Here $|0\ket$ and $|1\ket$ are the up and down eigenstate of the
$\sigma_z$, the photon polarization operator. If we measure the
state of a single photon, the Bell-state will collapse and the
state of the other particle will be completely determined if we
know the measurement result of the first photon. For example, if
we measure the state of photon A in the Bell state $\left\vert
\psi ^{-}\right\rangle $ and obtain $\left\vert
0\right\rangle $, then the state photon B will collapse to quantum state $%
\left\vert 1\right\rangle $.

In the QKD protocol in Ref.\cite{LL}, a set of $N$ ordered EPR
pairs, each randomly in one of the 4 Bell-states, is prepared and
divided into two sequences. Alice transmits the first sequence to
Bob, and then they measure a subset of photons in their hands
respectively. After that, they analyze the security of the
transmission for the first sequence. If they insure that the
channel  is safe, Alice sends the second sequence to Bob. Bob then
performs Bell-basis measurement on the ordered $N$ EPR pair to
read out the Bell-states. They perform a second eavesdropping
check. By analyzing error rate, they can ascertain whether they
have safely created a raw key or not. In this protocol, the
transmission is done in batches of $N$ EPR pairs. An advantage of
block-transmission protocol is that we can check the security of
the transmission  by measuring some of the photons in the first
step where Alice and Bob each holds a particle sequence in the
hands. Once the security of the quantum channel is ensured, which
means that an eavesdropper has not acquired the first particle
sequence, then no information will be leaked to her whatever she
may do to the second particle sequence.

 Because of this property,
this two-step QKD scheme can be modified for secure direct
communication. Here we first give the specific steps of the QSDC
protocol, they are

 (1) Alice and Bob agree on that each of the four Bell bases
can carry two-qubit classical information, and encode $\left\vert
\psi ^{-}\right\rangle $, $\left\vert \psi ^{+}\right\rangle $,
$\left\vert \phi ^{-}\right\rangle $ and $\left\vert \phi
^{+}\right\rangle $ as 00, 01, 10 and 11, respectively.

(2) Alice prepares an ordered $N$ EPR pairs in state $%
\left\vert \psi \right\rangle _{CM}=\left\vert \psi ^{-}\right\rangle =\frac{%
1}{\sqrt{2}}(\left\vert 0\right\rangle _{C}\left\vert
1\right\rangle _{M}+\left\vert 1\right\rangle _{C}\left\vert
0\right\rangle _{M})$. We
denote the $N$ ordered EPR pairs with [(P$_{1}$($C$),P$_{1}$($M$)), (P$_{2}$($%
C $),P$_{2}$($M$)), (P$_{3}$($C$),P$_{3}$($M$)), ... , (P$_{N}$($C$),P$_{N}$(%
$M $))]. Here the subscript indicates the pair order in the
sequence, C and M represent the two particles respectively.

(3) Alice takes one particle from each EPR pair to form an ordered
EPR partner particle sequence, say [P$_{1}$($C$), P$_{2}$($C$),
P$_{3}$($C$), ... , P$_{N}$($C$)]. It is called the checking
sequence or simply the C-sequence. The remaining EPR partner
particles compose another particle sequence [P$_{1}$($M$),
P$_{2}$($M$), P$_{3}$($M$), ... , P$_{N}$($M $)], and it is called
the message-coding sequence or the M-sequence for short.

(4) Alice sends the C-sequence [P$_{1}$($C$), P%
$_{2}$($C$), P$_{3}$($C$), ... , P$_{N}$($C$)] to Bob. Alice and
Bob then check eavesdropping by the following procedure: (a) Bob
chooses randomly a number of the photons from the C-sequence and
tell Alice which particles he has chosen. (b) Bob chooses randomly
one of the two sets of MBs, say $\sigma _{z}$ and $\sigma _{x}$\
to measure the chosen photons. (c) Bob tells Alice which MB he has
chosen for each photon and the outcomes of his measurements. (d)
Alice uses the same measuring basis as Bob to measure the
corresponding photons in the M-sequence and checks with the
results of Bob's. If no eavesdropping exists, their results should
be completely opposite, i.e., if Alice gets 0 (1), then Bob gets 1
(0). This is the first eavesdropping check.

After that, if the error rate is small, Alice and Bob can conclude
that there are no eavesdroppers in the line. Alice and Bob
continue to perform step 5; otherwise they have to discard their
transmission and abort the communication.

(5) Alice encodes her messages on the M-sequence and transmits it
to Bob. Before the transmission, Alice must encode the EPR pairs.
In order to guard for eavesdropping in this transmission, Alice
has to add a small trick in the M-sequence. She selects randomly
in the M-sequence some particles and perform on them randomly one
of the four operations. The number of such particles is not big as
long as it can provide an analysis of the error rate. Only Alice
knows the positions of these sampling particles and she keeps them
secret until the communication is completed. The remaining
M-sequence particles are used to carry the secret message
directly. To encode the message, we use the dense coding scheme of
Bennett and Wiesner\cite{dense} where the information is encoded
on an EPR pair with a local operation on a single qubit. Here we
generalize the dense coding idea into secure direct communication.
Different from dense coding,  in this protocol, the both particles
in an EPR are sent from Alice to Bob in two steps, and the
transmission of EPR pairs are done in block. Explicitly, Alice
makes one of the four unitary operations $(U_{0}$, $U_{1}$,
$U_{2}$ and $U_{3})$ to each of  her particles,
\begin{eqnarray}
U_{0}&&=I=\left\vert 0\right\rangle \left\langle 0\right\vert
+\left\vert 1\right\rangle \left\langle 1\right\vert ,
\label{O0}\\
 U_{1}&&=\sigma _{z}=\left\vert 0\right\rangle
\left\langle 0\right\vert -\left\vert 1\right\rangle \left\langle
1\right\vert ,  \label{O1}\\
 U_{2}&&=\sigma _{x}=\left\vert
1\right\rangle \left\langle 0\right\vert +\left\vert
0\right\rangle \left\langle 1\right\vert ,  \label{O2}\\
U_{3}&&=i\sigma _{y}=\left\vert 0\right\rangle \left\langle
1\right\vert -\left\vert 1\right\rangle \left\langle 0\right\vert
\label{O3}
\end{eqnarray}
and they transform the state $\left\vert \psi ^{-}\right\rangle $ into $%
\left\vert \psi ^{-}\right\rangle $, $\left\vert \psi ^{+}\right\rangle $, $%
\left\vert \phi ^{-}\right\rangle $, $\left\vert \phi
^{+}\right\rangle $ respectively. These operations correspond to
00, 01, 10, 11 respectively.

(6) After the transmission of M-sequence, Alice tells Bob the
positions of the sampling pairs and and the type of unitary
operations on them. Bob performs Bell-basis measurement on the C-
and M-sequences simultaneously. By checking the sampling pairs
that Alice has chosen, he will get an estimate of the error rate
in the M-sequence transmission. In fact, in the second
transmission, Eve can only disturb the transmission and cannot
steal the information because she can only get one particle from
an EPR pair.

(7) If the error rate of the sampling pairs is reasonably low,
Alice and Bob can then entrust the process, and continue to
correct the error in the secret message using error correction
method. Otherwise, Alice and Bob abandon the transmission and
repeat the procedures from the beginning.

(8) Alice and Bob do error correction on their results. This
procedure is exactly the same as that in QKD. However, to preserve
the integrity of the message, the bits preserving correction code,
like cascade\cite{CSS}, should be used.

As discussed above, Alice and Bob can ensure the security of the
C-sequence and Eve will be found out if she eavesdrops the quantum
line. It is of interest that Eve cannot read out the information
in the EPR pairs even if she captures one of the two sequences,
because no one can read the information from one particle of an
EPR pair alone. In this way, no secret message will be leaked to
Eve. It is secure. Moreover, the capacity is high in this
protocol, because each of EPR pair carries two bits of classical
information.

\section{SECURITY OF THE QSDC SCHEME}
\label{ss3}

Our QSDC protocol bases on EPR pair, so the proof of security is
similar to those in Refs.\cite{BF,IRV,WZY} with entangled photons.
The proof for the security of our QSDC protocol is based on the
security for the transmission of the C-sequence. If Alice and Bob
could not detect eavesdropper (Eve)in the transmission of the
C-sequence, Eve would capture easily the two photons in each EPR
pair and take Bell-basis measurement on them to read out the
secret message.

The transmission and the security check of the C-sequence in our
QSDC protocol is similar to the procedures in BBM92 QKD
protocol\cite{BBM}, where one particle in an EPR pair is sent to
Alice and the other is sent to Bob. Here the M-sequence particles
are retained securely in Alice's site. Before checking
eavesdropping, Eve has no access to the M-sequence particles.
Therefore the security of transmission for the C-sequence simply
reduces to the security of the BBM92 QKD protocol. The proof of
security for BM92 in ideal condition is done in Ref.\cite{IRV} and
that with practical conditions was given in detail in
Ref.\cite{WZY}. Hence our QSDC protocol is secure..

Now, let us give the reason that why we choose two sets of
measuring basis for checking the security of the transmission for
the C-sequence. According to Stinespring dilation theorem, as Eve
is limited only to eavesdropping on the quantum line between Alice
and Bob, her eavesdropping can be realized by a unitary operation,
say $\widehat{E}$ on a larger Hilbert space, $\left\vert
b,E\right\rangle \equiv \left\vert b\right\rangle _{B}\left\vert
E\right\rangle $. Then the state of composite system Alice, Bob
and Eve is
\begin{eqnarray}
\left\vert \psi \right\rangle =\sum\limits_{a,b\in
\{0,1\}}\left\vert \varepsilon _{a,b}\right\rangle \left\vert
a\right\rangle \left\vert b\right\rangle  \label{state1}
\end{eqnarray}
where $\left\vert \varepsilon _{a,b}\right\rangle $ describes
Eve's probe state, and $\left\vert a\right\rangle $ and
$\left\vert b\right\rangle $ are single photon states of Alice's
and Bob's in an each EPR pair, respectively. As in Ref.\cite{IRV},
the condition on the states of Eve's probe is

\begin{equation}
\sum\limits_{a,b\in \{0,1\}}\left\langle \varepsilon
_{a,b}|\varepsilon _{a,b}\right\rangle =1  \label{s1}
\end{equation}

As Eve can only eavesdrop the C-sequence before the first
checking, we can describe Eve's effect on the system as

\begin{eqnarray}
\left\vert 0,E\right\rangle \equiv \left\vert 0\right\rangle
_{B}\left\vert E\right\rangle &&=\alpha \left\vert 0\right\rangle
_{B}\left\vert \varepsilon _{00}\right\rangle +\beta \left\vert
1\right\rangle _{B}\left\vert \varepsilon _{01}\right\rangle
\equiv \alpha \left\vert 0,\varepsilon _{00}\right\rangle +\beta
\left\vert 1,\varepsilon _{01}\right\rangle, \label{state2}\\
\left\vert 1,E\right\rangle \equiv \left\vert 1\right\rangle
_{B}\left\vert E\right\rangle &&=\beta ^{^{\prime }}\left\vert
1\right\rangle _{B}\left\vert \varepsilon _{10}\right\rangle
+\alpha ^{^{\prime }}\left\vert 1\right\rangle _{B}\left\vert
\varepsilon _{11}\right\rangle \equiv \beta ^{^{\prime
}}\left\vert 0,\varepsilon _{10}\right\rangle +\alpha ^{^{\prime
}}\left\vert 1,\varepsilon _{11}\right\rangle,  \label{state3}
\end{eqnarray}

i.e, Eve's probe can be modelled by
\begin{equation}
\widehat{E}=\left(
\begin{array}{cc}
\alpha & \beta ^{^{\prime }} \\
\beta & \alpha ^{^{\prime }}%
\end{array}%
\right).  \label{O4}
\end{equation}

Since $\widehat{E}$ has to be\ unitary, the complex numbers\ $\alpha $%
,\ $\beta $,\ $\alpha ^{^{\prime }}$\ and $\beta ^{^{\prime }}$\
must satisfy

\begin{eqnarray}
\left\vert \alpha \right\vert ^{2}+\left\vert \beta ^{^{\prime
}}\right\vert
^{2} &=&1,  \label{s2} \\
\left\vert \alpha ^{^{\prime }}\right\vert ^{2}+\left\vert \beta
\right\vert
^{2} &=&1, \nonumber \\
\alpha \beta ^{\ast }+\alpha ^{^{\prime }\ast }\beta ^{^{\prime }}
&=&0. \nonumber
\end{eqnarray}

We get the following relations

\begin{equation}
\left\vert \beta ^{^{\prime }}\right\vert ^{2}=\left\vert \beta
\right\vert ^{2},\left\vert \alpha ^{^{\prime }}\right\vert
^{2}=\left\vert \alpha \right\vert ^{2}.  \label{s3}
\end{equation}

For Alice and Bob, the action of Eve's eavesdropping will
introduce an error
rate%
\begin{equation}
\epsilon =\left\vert \beta \right\vert ^{2}=\left\vert \beta
^{^{\prime }}\right\vert ^{2}=1-|\alpha |^{2}=1-|\alpha ^{^{\prime
}}|^{2}  \label{s4}
\end{equation}

If Eve can only capture one photon in each EPR pair, she gets no
information. The way Eve can steal information is that she pretend
Bob to receive the C-sequence and send a fake sequence to Bob. If
Alice and Bob could not find out her action, Eve would intercept
the M-sequence and read out the information in the EPR pairs. That
is to say, only when Alice and Bob ascertain that there is no
eavesdropper monitoring the quantum line, they transmit the
M-sequence. We can calculate the information Eve can maximally
gain. When the C-sequence particles reach Bob, its reduced density
matrix is
\begin{equation}
\rho _{B}=Tr_{A}(\rho _{AB})=Tr_{A}(\left\vert \psi \right\rangle
_{ABAB}\left\langle \psi \right\vert )=\frac{1}{2}\left(
\begin{array}{cc}
1 & 0 \\
0 & 1%
\end{array}%
\right),   \label{O5}
\end{equation}%
that is to say, Bob's photon can be in either state $\left\vert
0\right\rangle $ or $\left\vert 1\right\rangle $ with equal probability $P=%
\frac{1}{2}$.

Similar to that in Ref.\cite{BF}, first let us suppose that the
quantum state of the photon in the hand of Alice is $\left\vert
0\right\rangle $, i.e., Alice takes measurement on the photon in
her hand with single photon detector and the state is $\left\vert
0\right\rangle $. Then the state of the system composed of Bob's
photon and Eve's probe can be described by

\begin{equation}
\left\vert \psi ^{^{\prime }}\right\rangle =\widehat{E}\left\vert
0,E\right\rangle \equiv \widehat{E}\left\vert 0\right\rangle
_{B}\left\vert E\right\rangle =\alpha \left\vert 0\right\rangle
_{B}\left\vert \varepsilon _{00}\right\rangle +\beta \left\vert
1\right\rangle _{B}\left\vert \varepsilon _{01}\right\rangle
\equiv \alpha \left\vert 0,\varepsilon _{00}\right\rangle +\beta
\left\vert 1,\varepsilon _{01}\right\rangle ,\label{state4}
\end{equation}

\begin{equation}
\rho ^{^{\prime }} =\left\vert \alpha \right\vert ^{2}\left\vert
0,\varepsilon _{00}\right\rangle \left\langle 0,\varepsilon
_{00}\right\vert +\left\vert \beta \right\vert ^{2}\left\vert
1,\varepsilon _{01}\right\rangle \left\langle 1,\varepsilon
_{01}\right\vert +\alpha \beta ^{\ast }\left\vert 0,\varepsilon
_{00}\right\rangle \left\langle 1,\varepsilon _{01}\right\vert
+\alpha ^{\ast }\beta \left\vert 1,\varepsilon _{01}\right\rangle
\left\langle 0,\varepsilon _{00}\right\vert. \label{O6}
\end{equation}

After encoding of the unitary operations $U_{0}$, $U_{1}$, $U_{2}$
and $U_{3}$ with the probabilities $p_{0}$, $p_{1}$, $p_{2}$ and
$p_{3}$ respectively, the state reads

\begin{eqnarray}
\rho ^{^{^{\prime \prime }}} &=&(p_{0}+p_{3})\left\vert \alpha
\right\vert ^{2}\left\vert 0,\varepsilon _{00}\right\rangle
\left\langle 0,\varepsilon _{00}\right\vert
+(p_{0}+p_{3})\left\vert \beta \right\vert ^{2}\left\vert
1,\varepsilon _{01}\right\rangle \left\langle 1,\varepsilon
_{01}\right\vert
\label{O7} \\
&&+(p_{0}-p_{3})\alpha \beta ^{\ast }\left\vert 0,\varepsilon
_{00}\right\rangle \left\langle 1,\varepsilon _{01}\right\vert
+(p_{0}-p_{3})\alpha ^{\ast }\beta \left\vert 1,\varepsilon
_{01}\right\rangle \left\langle 0,\varepsilon _{00}\right\vert  \nonumber \\
&&+(p_{1}+p_{2})\left\vert \alpha \right\vert ^{2}\left\vert
1,\varepsilon _{00}\right\rangle \left\langle 1,\varepsilon
_{00}\right\vert +(p_{1}+p_{2})\left\vert \beta \right\vert
^{2}\left\vert 0,\varepsilon
_{01}\right\rangle \left\langle 0,\varepsilon _{01}\right\vert  \nonumber \\
&&+(p_{1}-p_{2})\alpha \beta ^{\ast }\left\vert 1,\varepsilon
_{00}\right\rangle \left\langle 0,\varepsilon _{01}\right\vert
+(p_{1}-p_{2})\alpha ^{\ast }\beta \left\vert 0,\varepsilon
_{01}\right\rangle \left\langle 1,\varepsilon _{00}\right\vert
\nonumber
\end{eqnarray}

\bigskip which can be rewritten in the orthogonal basis $\{\left\vert
0,\varepsilon _{00}\right\rangle ,\left\vert 1,\varepsilon
_{01}\right\rangle ,\left\vert 1,\varepsilon _{00}\right\rangle
,\left\vert
0,\varepsilon _{01}\right\rangle \}$,%
\begin{equation}
\rho ^{^{^{\prime \prime }}}=\left(
\begin{array}{cccc}
(p_{0}+p_{3})\left\vert \alpha \right\vert ^{2} &
(p_{0}-p_{3})\alpha \beta
^{\ast } & 0 & 0 \\
(p_{0}-p_{3})\alpha ^{\ast }\beta & (p_{0}+p_{3})\left\vert \beta
\right\vert ^{2} & 0 & 0 \\
0 & 0 & (p_{1}+p_{2})\left\vert \alpha \right\vert ^{2} &
(p_{1}-p_{2})\alpha \beta ^{\ast } \\
0 & 0 & (p_{1}-p_{2})\alpha ^{\ast }\beta &
(p_{1}+p_{2})\left\vert \beta
\right\vert ^{2}%
\end{array}%
\right)  \label{O8}
\end{equation}

where $p_{0}+p_{1}+p_{2}+p_{3}=1$.

The information $I_{0}$\ that Eve can get is equal to the Von
Neumann entropy, i.e,

\begin{equation}
I_{0}=\sum\limits_{i=0}^{3}-\lambda _{i}\log _{2}\lambda _{i}
\label{information}
\end{equation}

where $\lambda _{i}$ (i=0,1,2,3) are the eigenvalues of $\rho
^{^{^{\prime \prime }}}$, which are

\begin{eqnarray}
\lambda _{0,1} &=&\frac{1}{2}(p_{0}+p_{3})\pm \frac{1}{2}\sqrt{%
(p_{0}+p_{3})^{2}-16p_{0}p_{3}|\alpha |^{2}|\beta |^{2}}  \label{root1} \\
&=&\frac{1}{2}(p_{0}+p_{3})\pm \frac{1}{2}\sqrt{%
(p_{0}+p_{3})^{2}-16p_{0}p_{3}(\epsilon -\epsilon ^{2})},
\nonumber
\end{eqnarray}

\begin{eqnarray}
\lambda _{2,3} &=&\frac{1}{2}(p_{1}+p_{2})\pm \frac{1}{2}\sqrt{%
(p_{1}+p_{2})^{2}-16p_{1}p_{2}|\alpha |^{2}|\beta |^{2}}  \label{root2} \\
&=&\frac{1}{2}(p_{1}+p_{2})\pm \frac{1}{2}\sqrt{%
(p_{1}+p_{2})^{2}-16p_{1}p_{2}(\epsilon -\epsilon ^{2})}.
\nonumber
\end{eqnarray}
If the four operations distribute with equal probability, that is $%
p_{0}=p_{1}=p_{2}=p_{3}=\frac{1}{4}$, Eve can get 1 bit of
information from each EPR pair with the error rate $\epsilon =0$.
In fact, the simple way for Eve to steal information is that Eve
measures each photon with MB $\sigma _{z}$ and Alice and Bob
cannot find out the action of Eve's. Even though Eve cannot read
out the information of phase in EPR pair, she can distinguish the
value of each bit. That is to say, she can distinguish the
operations \{$U_{0}$,$U_{1}$\} from \{$U_{2}$, $U_{3}$\}. This is
an intrinsic limitation on the coding in Ref.\cite{BF}.

Surely the proof and the above discussion are based on ideal
condition and do not take into account noise in transmission. In
fact, in low noise channel, the photon loss will be small, and
Eve's action will increase either the error rate or loss of
signal, so the security of C-sequence is assured if Alice and Bob
do the first eavesdropping check and analyze the error rate and
the  efficiency. On the contrary, if quantum channel loss is
sufficiently high, two problems arise. The first one is the
security of transmission of the C-sequence which requires Alice
and Bob share a sequence of entangled states securely. The other
is the loss of the M-sequence. Without measurement Bob cannot make
sure whether he receives the particles or not in the C-sequence
and Alice must encode all particles in M-sequence. In this way,
Eve's eavesdropping cannot be detected if she captures some of
particles in C-sequence and sends the others to Bob with a better
quantum channel with which the lossy efficiency of all the photons
is not increased. Eve intercepts the M-sequence and do Bell-basis
measurement and then gets some of the secret message. This is the
danger of not sharing a sequence of EPR pairs securely. In order
to avoid the attack on C-sequence and share a sequence of EPR
pairs securely, Bob can perform quantum entanglement
swapping\cite{Pan} on the particles he receives first and then
gets a subset of C-sequence of particles entangled with
Alice's(called the C'-sequence): if there are indeed particles
there, the swapping will succeed otherwise the swapping will fail.
The swapping operation here serves as a particle existence
detection. Then Bob chooses randomly a subset of the C'-sequence
particles and measures them with either $\sigma _{z} $ or $\sigma
_{x}$. Alice only encodes on the subset of M-sequence
corresponding to
the sub-C-sequence(called the M'-sequence) on which Bob succeeds in
 quantum entanglement swapping.
 With these two procedures Alice and Bob can share a sub-sequence
of EPR pairs truly and the action of Eve's can be detected even in
a highly lossy quantum channel. In practical applications, some
coding using redundancy is necessary as has been extensively in
classical communications. For instance,  several bits mayve used
to code a single bit for instance using the CSS coding
method\cite{CSS}. In this way, Alice and Bob must pay a lot of
source for the correlated results.

\section{IMPLEMENTATION ISSUES}
\label{ss4}

In our scheme, we need to store the checking sequence of photons
for a while, to make eavesdropping check and wait for the
M-sequence of photons to come. This is the price to pay for the
improved security and enhanced efficiency. Here we propose two
ways to realize this. One is using light storage device, and the
other is to use optical delays.

It has already been demonstrated experimentally that light can be
stored together with their quantum states\cite{Phillips,Liu}. With
the electromagnetic induced transparency technique, the C-sequence
of photons can be stored for a while to complete the eavesdropping
check and the travelling of the M-sequence. At present, the
technique may not be mature enough for a practical implementation
of the proposed QSDC scheme. However, as it may be the only light
storage device, together with its roles in quantum computation, it
is extremely demanding that this technique be developed further.

Another realization is to use optical delays. This is a well
developed technology and is experimentally feasible. Instead of
producing an ordered EPR pairs in space at the same time, we can
produce a time ordered EPR pair sequence. As shown in Fig.2, a
sequence of EPR pairs are produced at Alice's site. One after
another photon in the C-sequence is sent to Bob's site through the
upper line first. The corresponding M-sequence is sent to Bob
through the lower transmission line. However, the M-sequence is
delayed by $\tau $ at Alice's site before it enters the insecure
channel. When the C-sequence reaches Bob, Bob selects randomly
some photons for eavesdropping check. He
measures those chosen photons randomly in the $\sigma _{x}$\ or $\sigma _{z}$%
\ basis and he announces publicly the positions, and the
measuring-basis and the outcomes of the measurement for these
chosen photons. After hearing these results, Alice performs
measurement using the same measuring-basis as Bob's on the
corresponding photons in the M-sequence. If the error rate is
below a predetermined threshold, she concludes that the quantum
channel is secure and preforms coding unitary operation on the
M-sequence particles. During the M-sequence transmission, some
randomly chosen photons are used to check the transmission error
rate. In these chosen sampling photons, an operation randomly
chosen form the four operations is applied. Therefore after Bob
receives the sampling pairs and combines with his  partner photons
in the C-sequence, he can recover these operations using
Bell-basis measurement. These sampling pairs will give an error
rate estimate of the second transmission, and this error rate will
be used as a parameter later in error correction process.

A very important quantity is the delay $\tau $. It depends on the
distance between Alice and Bob, the number N in each block, and
the number of photons transmitted per unit time, $f$. For
simplicity, we ignore the times it takes
for the eavesdropping check measurement, and the coding operation. Then $%
\tau $ must be long enough for a photon to travel to Bob, and Bob
makes measurement and tells Alice the result, and then sends the
M-sequence particles to Bob. Thus it must be 3 times of the period
for a photon to travel from Alice to Bob. If this has to be done
for a block of N pairs, additional time $\frac{N}{f}$ has to be
added. Thus the delay should be

\begin{equation}
\tau \geq \frac{3L}{c}+\frac{N}{f}  \label{time}
\end{equation}
where L is the distance between Alice and Bob, and $c$ is the
velocity of light in quantum channel. Complete Bell-basis
measurement is also highly demanding, and has been demonstrated
recently\cite{Kim}.

\section{Discussion and Summary}
\label{ss5}

The presented scheme resembles more to a quantum key distribution
protocol. In fact, after Bob receives the checking sequence, Alice
and Bob can establish a common one-time-pad key by measuring their
particles using a randomly chosen basis from the $\sigma_z$ or
$\sigma_x$ basis, which is a variant of the Ekert91\cite{Ekert}
QKD and the BBM92\cite{BBM} QKD protocol. Then the secret message
can be encoded using this one-time-pad key and transmitted through
a classical channel. The important distinction between quantum
direct communication and the quantum key distribution scheme is
that in the quantum direct communication scheme no classical key
is ever established, but rather an inherently quantum mechanical
resource (the shared EPR pairs) takes over the role of the key.
With the development of efficient EPR source and Bell-state
measurement device, quantum direct communication may become easier
to realize and be favored in some specific applications.

In summary, a novel QSDC scheme is proposed and secret messages
can be coded directly over a quantum channel with security. In
this scheme, A block of entangled particles is divided into two
sequences, the checking sequence and message-coding sequence. They
are sent from Alice to Bob in two steps. The security is assured
by the secure transmission of the checking sequence. Moreover, the
scheme makes full use of the 2-qubit in an EPR pair. We also
propose concrete experimental setup for its realization. The
scheme is completely secure for an ideal noiseless channel, and it
conditionally secure with a noisy channel.

\section*{ACKNOWLEDGMENT}

This work is supported the National Fundamental Research Program
Grant No. 001CB309308, China National Natural Science Foundation
Grant No. 60073009, the Hang-Tian Science Fund, and the Excellent
Young University Teachers' Fund of Education Ministry of China.

\bigskip

\begin{figure}
\begin{center}
\caption{ Illustration of the QSDC protocol.  Alice prepares the
ordered $N$ EPR pair in the same quantum states and divides them
into two partner-particle sequences.  She firstly sends one
sequence to Bob for checking eavesdropping by choosing a fraction
of particles to measure with randomly chosen measuring basis.  If
the quantum line is secure, Alice encodes the partner EPR pairs
using four unitary operations the secret messages and sends the
second sequence to Bob.}
\includegraphics[width=16cm,angle=0]{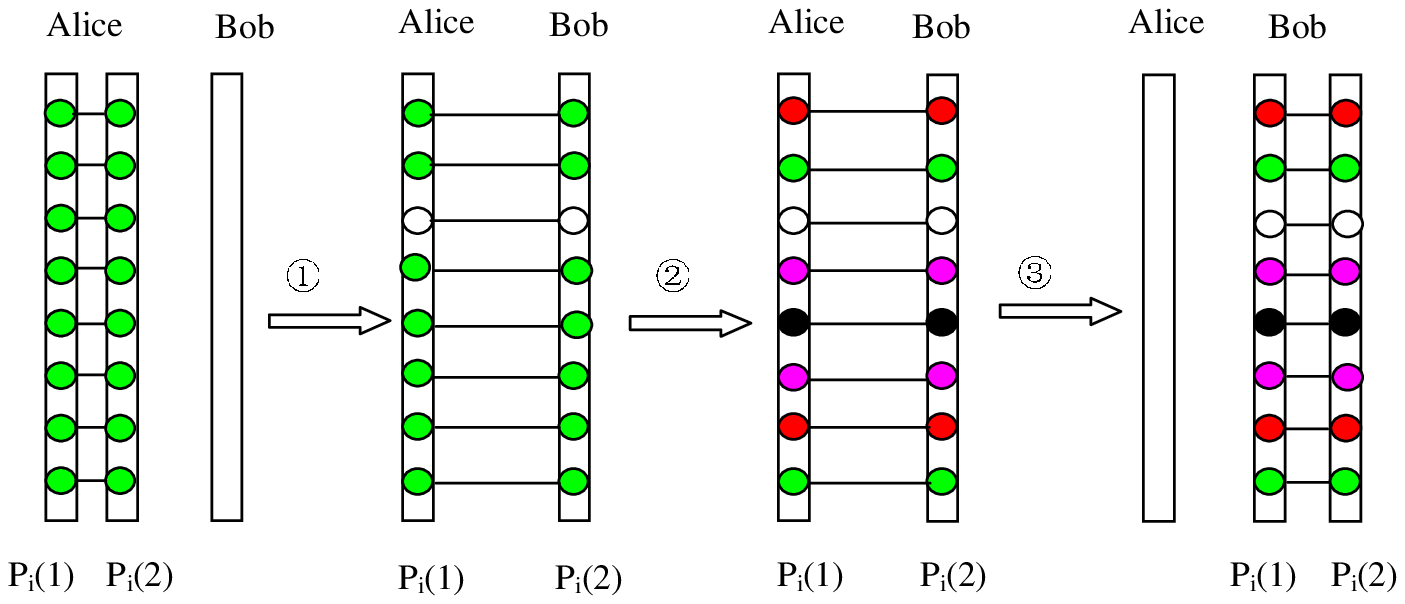} \label{f1}
\end{center}
\end{figure}

\begin{figure}
\begin{center}
\caption{ An example for the QSDC scheme using optical delays.
SR1, SR2, SR3 and SR4 represent optical delay; CE1 and CE2 are
used to describe the procedure for checking eavesdropping; CM is
coding message sequence according to secret message.}
\includegraphics[width=8cm,angle=-90]{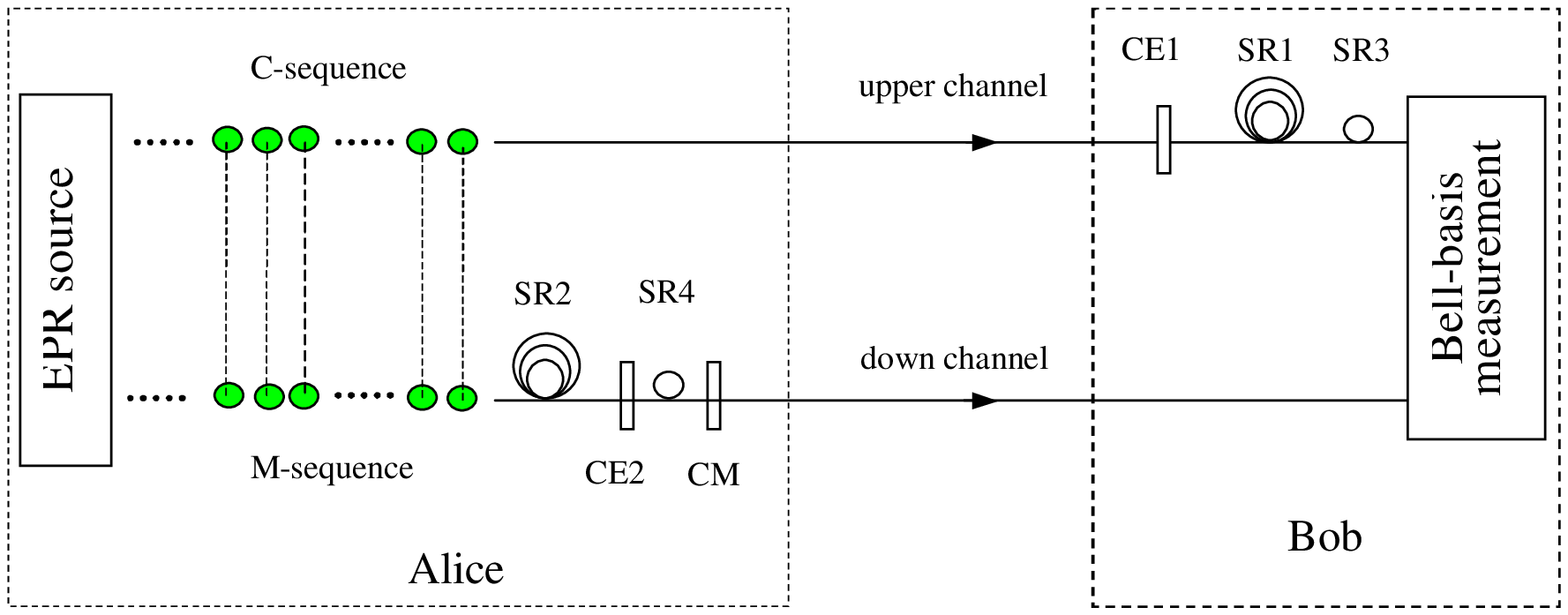} \label{f2}
\end{center}
\end{figure}

\end{document}